\RequirePackage{fix-cm}
\documentclass[twocolumn]{svjour3}          % twocolumn
\smartqed  % flush right qed marks, e.g. at end of proof
\usepackage{amsmath,bm}
\usepackage{graphicx}
\usepackage{amsfonts}
\usepackage{algorithm}
\usepackage[noend]{algpseudocode}
\usepackage[para]{threeparttable}
\usepackage{booktabs} % To thicken table lines
\usepackage{boldline}
\usepackage[justification=raggedright,singlelinecheck=false,font=small]{caption}
\usepackage{xcolor}
\usepackage{cite}
\usepackage{url}
\usepackage{multirow}
\usepackage{dblfloatfix}
\usepackage[square,numbers,sort&compress]{natbib}
\usepackage{nicefrac} % For comparison
\usepackage[switch]{lineno}

\newcommand{\p}{\mathbf{p}}
\newcommand{\rr}{\mathbf{r}}
\newcommand{\A}{\mathbf{A}}
\newcommand{\B}{\mathbf{B}}
\newcommand{\M}{\mathbf{M}}
\newcommand{\D}{\mathbf{D}}

\newcommand{\x}{\mathbf{x}}

\newcommand{\y}{\mathbf{y}}

\newcommand{\Vvarepsilon}{\bm{\varepsilon}}
\newcommand{\etal}{\textit{et al}.~}

\begin{document}\sloppy

\title{FL-MISR: Fast Large-Scale Multi-Image Super-Resolution for Computed Tomography Based on Multi-GPU Acceleration
}

\titlerunning{FL-MISR: Fast Large-Scale Multi-Image Super-Resolution}        % if too long for running head

\author{Kaicong Sun   \and
        Trung-Hieu Tran \and
        Jajnabalkya Guhathakurta \and
        Sven Simon      
}


\institute{Kaicong Sun, Trung-Hieu Tran, Jajnabalkya Guhathakurta, Sven Simon  \at Institute of Parallel and Distributed Systems, University of Stuttgart, 70569, Stuttgart, Germany\\
              %Tel.: +49-711-68588421\\
              \email{kaicong.sun@ipvs.uni-stuttgart.de} \\
              \email{trung-hieu.tran@ipvs.uni-stuttgart.de} \\
              \email{jajnabalkya.guhathakurta@ipvs.uni-stuttgart.de} \\
              \email{sven.simon@ipvs.uni-stuttgart.de} \\   
}

\date{Received: date / Accepted: date}
%\linenumbers

\maketitle

\begin{abstract}
Multi-image super-resolution (MISR) usually outperforms single-image super-resolution (SISR) under a proper inter-image alignment by explicitly exploiting the inter-image correlation. However, the large computational demand encumbers the deployment of MISR in practice. In this work, we propose a distributed optimization framework based on data parallelism for fast large-scale MISR using multi-GPU acceleration named FL-MISR. The scaled conjugate gradient (SCG) algorithm is applied to the distributed subfunctions and the local SCG variables are communicated to synchronize the convergence rate over multi-GPU systems towards a consistent convergence. Furthermore, an inner-outer border exchange scheme is performed to obviate the border effect between neighboring GPUs. %Inter-GPU communication for the exchange of local variables and overlapped regions is enabled to impose a consensus convergence of the distributed task allocated to each GPU node. 
The proposed FL-MISR is applied to the computed tomography (CT) system by super-resolving the projections acquired by subpixel detector shift. The SR reconstruction is performed on the fly during the CT acquisition such that no additional computation time is introduced. %We evaluated FL-MISR quantitatively and qualitatively on multiple objects including aluminium cylindrical phantoms, QRM bar pattern phantoms, and concrete joints. 
FL-MISR is extensively evaluated from different aspects and experimental results demonstrate that FL-MISR effectively improves the spatial resolution of CT systems in modulation transfer function (MTF) and visual perception. Comparing to a multi-core CPU implementation, FL-MISR achieves a more than 50$\times$ speedup on an off-the-shelf 4-GPU system. %Last but not least, FL-MISR provides also very promising performance for natural images.  
\end{abstract}

\section{Introduction}
\label{sec:introduction}
Super-resolution (SR) is a fundamental task in image processing and has been an attractive research field for decades~\cite{SRReview1, SRReview2, SRReview3}. SR is an algorithm-based image enhancement technique dedicated to improving the spatial resolution of the imaging systems beyond the hardware limit by exploiting the low-resolution (LR) acquisitions and it is widely applied in many applications such as medical diagnostics, surveillance, and remote sensing. 

In recent years, we have witnessed tremendous progress of deep learning in multiple image processing and computer vision tasks such as image denoising~\cite{DnCNN, KPN}, super-resolution~\cite{SRCNN, EDSR}, deformable registration~\cite{VoxelMorph, FDRN}, and semantic segmentation~\cite{nnUNet, unet++}. Despite of the great success of deep learning in SR, most of the work focuses on single-image SR (SISR)~\cite{SRCNN, FSRCNN, VDSR, DPSR, EDSR, ESRGAN, ReCNN, mDCSRN}. In fact, SR reconstruction can significantly benefit from the available correlated input images which are captured of the same view. Multi-image SR (MISR) exploits the correspondences entailed in the multiple input images and usually outperforms SISR when the relative movements between the reference image and the other input images are well estimated. However, the learning-based MISR approaches in the literature are mainly dedicated to video applications~\cite{VSRnet, ESPCN, RBPN, FRVSR, ERVSR}. Besides, the quality of the learning-based methods highly depends on the fidelity of the training datasets. In practice, preparing synthetic datasets which adequately resemble the real-world measurements covering diverse imaging conditions would be challenging. Furthermore, although learning-based approaches are able to describe more sophisticated image priors, ``hallucinated'' structures can be unpredictably constructed which may impede the employment of the trained models in applications such as metrology and quality control.

Different from the deep learning-based SR methods, optimization-based MISR algorithms~\cite{BTV, Yue, IRWSR, MPG, MPGBSWTV} reconstruct the latent high-resolution (HR) image explicitly based on the real acquisitions but not the training datasets. Nowadays, due to the technological development of sensor manufacturing, sensors or detectors with large resolutions such as 8, 16 Mpixels or even higher are employed in applications such as medical imaging and industrial inspection. Coping with large-scale multi-image input can be computationally expensive and hardware costly. The optimization-based SR methods usually suffer from the iterative manner which leads to undesirable computation time. In this work, we present a multi-GPU accelerated framework for large-scale MISR reconstruction based on distributed optimization. The proposed framework is applied to the computed tomography (CT) imaging system and achieves a real-time SR reconstruction during the CT acquisition without introducing additional computation time. The contribution of this work can be summarized as following:
\begin{itemize}
\item We propose a distributed optimization framework for MISR, named FL-MISR, dealing with large sized images based on multi-GPU acceleration. Each GPU accounts for an allocated partition and the latent SR image is obtained by image fusion. %Inter-GPU communication is performed to preserve a consensus convergence rate and avoid border effects.
\item In order to obtain a consistent resolution enhancement among all the GPUs, the update of the partitions is synchronized by unifying the local variables of the scaled conjugate gradient (SCG) method. To avoid border effect between neighboring GPUs, an inner-outer border exchange scheme is performed. 
\item The proposed FL-MISR is applied to real-time CT imaging by super-resolving the projections acquired via subpixel detector shift. Extensive evaluation from different aspects demonstrates that FL-MISR not only achieves a significant resolution enhancement for CT systems but also provides very promising results for natural images. Comparing to a multi-core CPU implementation, FL-MISR achieves a more than $50\times$ speedup on a 4-GPU system. 
\end{itemize}

%It should be noted that the proposed multi-GPU MISR reconstruction framework composes a generalized decomposition scheme for handeling large images and it is not confined to CT.

\section{Related Work}
\label{sec:related works}
\subsection{Optimization-Based Iterative Methods}
In the literature, conventional optimization-based iterative SR methods can be traced back to 1980's and they are mainly grouped into two categories: the frequency domain based and the spatial domain based methods~\cite{SRReview1, SRReview2}. In~\cite{Huang}, Huang~\etal firstly address the MISR problem in the frequency domain. Although the frequency domain based methods have low computational complexity, they behave extremely sensitive to model errors and have limited ability to integrate a priori knowledge as regularization. The majority of the iterative MISR approaches solve the problem in the spatial domain based on the maximum likelihood (ML), the maximum a posteriori (MAP), and the projection onto convex sets (POCS)~\cite{Stark,Elad,Tipping,BTV,Jens,IRWSR,Xu,MPGBSWTV}. Most of the work focuses on the reconstruction accuracy and only few concerns the performance in computation time. Specially, Elad~\etal\cite{Elad} propose a fast MISR algorithm concerning the special case of pure translation and space invariant blur. In~\cite{BTV}, Farsiu~\etal present a robust MISR method based on MAP using the L1 norm data fidelity term and the bilateral total variation regularization. Jens~\etal\cite{Jens} introduce a GPU-accelerated MISR approach for image-guided surgery which supports a 2$\times$ SR reconstruction from 4 LR images of size 200$\times$200 in 60 $ms$. However, due to the GPU memory limit, their method can not handle large sized images. In~\cite{Anger}, the authors propose a fast MISR method which composes of registration, fusion, and sharpening for satellite images using high-order spline interpolation. Nevertheless, purely image fusion is performed on a GPU and the rest two steps are on the CPU which results in a degraded performance in runtime. 
    
\subsection{Deep Learning-Based Methods}
In the last decade, deep learning has been very successfully adopted in SR and has harvested fruitful results. Dong~\etal\cite{SRCNN} introduce the convolutional neural network (CNN) into SISR which demonstrates the great potential of CNN for feature extraction. Inspired by the distinguished performance of CNN, a series of work from plain CNN to densely connected GAN, from 2D natural image to 3D medical volume, has been successively proposed~\cite{FSRCNN,VDSR,EDSR,ReCNN,ESRGAN,mDCSRN}. Comparing to the traditional iterative methods, CNN-based SR approaches focus on super-resolving single LR image by exploiting the relation learned exclusively from the LR-HR image pairs in the external example database. The learning-based MISR methods are mainly proposed to cope with natural video streams~\cite{VSRnet,ESPCN,RBPN,FRVSR}. Although some work is intended for real-time applications using GPU or FPGA~\cite{FRVSR,Kim,ERVSR}, the video SR (VSR) performance highly relies on the fidelity of the synthesized LR-HR frame pairs and the quality of the training datasets. Furthermore, since the supervised learning scheme requires the ground truth (GT) HR images during the training phase, the performance of the trained model will be limited by the available quality of the GT acquired in practice. It is especially true for CT imaging due to the lack of publicly available high-quality HR datasets like DIV8K~\cite{DIV8K} for natural images. 

To the best of our knowledge, the literature on GPU-accelerated MISR methods for large-scale images is very limited despite of its importance. In this paper, we extend our previous work~\cite{iCT} mainly in two folds. First, the locally applied scaled conjugate gradient (SCG) algorithm is adapted to achieve a synchronized convergence rate over multi-GPU systems. Second, instead of performing region averaging, we employ the so-called inner-outer border exchange scheme to preserve the sharpness of the overlapped regions. Particularly, in~\cite{iCT} we introduce a multi-GPU implementation of a MISR method based on data parallelism where each GPU deals with an allocated partition of the latent SR image. Although overlapped regions between neighboring GPUs are exchanged and averaged, we found that the resolved SR image in~\cite{iCT} is not globally optimized and the fused SR image suffers from inhomogeneous resolution enhancement due to the inconsistent convergence rate of the local SCG and region averaging. %To address these issues, in this work we adapt the SCG algorithm towards a consensus convergence for distributed optimization over multi-GPUs. 
We address these issues in this work and propose a generalized framework for multi-GPU supported MISR. We conducted extensive experiments to validate the proposed FL-MISR. Experimental results show that the exchange of local SCG variables and overlapped regions among GPUs has limited impact on the overall performance of runtime and leads to a consensus convergence over multi-GPUs without causing border effects. Besides, it is shown that super-resolving four input images of size 4096$\times$4096 by an upscaling of 2$\times$ can be achieved within 2.4$s$ on a 4-GPU system. 

%To the best of our knowledge, the literature on GPU-accelerated MISR methods for large-scale images is very limited despite of its importance. In this paper, we extend our previous work~\cite{iCT} by adapting the employed scaled conjugate gradient (SCG) algorithm to achieve global optimum for distributed reconstruction of MISR on multi-GPU systems. Particularly, in~\cite{iCT} we introduce a multi-GPU implementation of a MISR method based on data parallelism where each GPU deals with an allocated partition of the latent SR image. Although overlapped regions between neighboring GPUs are exchanged to avoid border discontinuity, we found that the resolved SR image in~\cite{iCT} is not globally optimized and the fused SR image suffers from inhomogeneous resolution due to the inconsistent local variables of the SCG algorithm. To address this issue, in this work we adapt the SCG algorithm towards a consensus convergence for distributed optimization over multi-GPUs. We conducted extensive experiments to validate the propose FL-MISR. Experimental results show that the exchange of local SCG variables and overlapped regions among GPUs has limited impact on the overall performance of runtime and leads to a consensus convergence over multi-GPUs without causing border effects. Besides, it is shown that super-resolving four input images of size 4096$\times$4096 by an upscaling of 2$\times$ can be achieved within 2.4$s$ on a 4-GPU system.  

\section{Methods}
\label{sec:method}	
\subsection{Distributed Optimization for MISR}\label{subsec:distributed optimization}
The common formulation of SR model in the pixel domain is presented as 
\begin{linenomath}
\begin{equation}\label{eq:sisr}
\y = \A\x+\Vvarepsilon(\x)
\end{equation}
\end{linenomath}
with $\x\in\mathbb{R}^{n\times 1}, \y\in\mathbb{R}^{m\times 1}$ being respectively the latent and captured image rearranged in lexicographic order. The system matrix $\A\in\mathbb{R}^{m\times n}$ is usually expressed as $\A=\D\B\M$ with $\D\in\mathbb{R}^{m\times n}, \B\in\mathbb{R}^{n\times n}$, and $\M\in\mathbb{R}^{n\times n}$ describing the decimation, blurring, and motion effects, respectively. The vector $\Vvarepsilon(\x)\in\mathbb{R}^{m\times 1}$ denotes the additive noise existing in the imaging systems. More detailed description of the system model can be found in~\cite{MPGBSWTV}. To simplify the calculation, in this work we assume $\Vvarepsilon(\x)$ is an intensity-independent additive noise and the system matrix $\A$ is known.

Since SR is an ill-posed problem, involving a well-defined image prior can effectively constrain the solution domain. Therefore, MAP estimator is preferably adopted for SR reconstruction. The posterior probability $P(\x|\y)$ of the SR image $\x$ is formulated based on the Bayes' theorem:
\begin{linenomath}
\begin{equation}\label{eq:bayes}
P(\x|\y) = \frac{P(\y|\x)P(\x)}{P(\y)}.
\end{equation}
\end{linenomath}
Assuming the noise $\varepsilon_i\in\Vvarepsilon$ in each pixel $i$ is white Gaussian and i.i.d with $\varepsilon_i\sim N(0, \sigma^2)$ and $P(\varepsilon_i)=\frac{1}{\sqrt{2\pi\sigma^2}}e^{-\frac{\varepsilon_i^2}{2\sigma^2}}$, we yield the likelihood function as
\begin{linenomath}
\begin{equation}\label{eq:likelihood}
P(\y|\x) = \prod\limits_{i=1}^{m}P(y_i|\x) = \left(\frac{1}{\sqrt{2\pi\sigma^2}}\right)^m e^{-\frac{||\A\x-\y||^2_2}{2\sigma^2}}
\end{equation}
\end{linenomath}
Taking the natural logarithm, the associated negative log-likelihood can be formulated as
\begin{linenomath}
\begin{equation}\label{eq:log}
-log\left(P(\y|\x)\right) = \frac{1}{2\sigma^2}||\A\x-\y||^2_2+c
\end{equation}
\end{linenomath}
where $c$ is a constant. For brevity, we will omit the weight $\frac{1}{2\sigma^2}$ and the constant $c$ in the latter formulation. 

\iffalse
In case of additive white Laplacian noise which models the impulse noise (Salt \& Pepper noise), the likelihood function becomes
\begin{equation}\label{eq:likelihoodLaplacian}
P(\y|\x) = \prod\limits_{i=1}^{m}P(y_i|\x) = \left(\frac{1}{2\sigma}\right)^m e^{-\frac{||\A\x-\y||_1}{\sigma}}
\end{equation}
and the data fidelity term has the form
\begin{equation}\label{eq:logLaplacian}
-log\left(P(\y|\x)\right) = \frac{1}{\sigma}||\A\x-\y||_1+c
\end{equation}
\fi
For MISR with $k$ independent LR images $\y_i$, $i\in[1\dots k]$, the posterior probability can be extended as
\begin{linenomath}
\begin{equation}\label{eq:bayes}
P(\x|\y_1\dots\y_k) = \frac{P(\y_1\dots\y_k|\x)P(\x)}{\prod\limits_{i=1}^{k}P(\y_i)} = \frac{\prod\limits_{i=1}^{k}P(\y_i|\x)P(\x)}{\prod\limits_{i=1}^{k}P(\y_i)}
\end{equation}
\end{linenomath}
and the data fidelity term is hence formulated by
\begin{linenomath}
\begin{equation}\label{eq:l2misr}
-log\left(\prod\limits_{i=1}^{k}P(\y_i|\x)\right) = \sum\limits_{i=1}^{k}||\A_i\x-\y_i||^2_2.
\end{equation}
\end{linenomath}
%where $i$ indicates the index of the captured LR image. 
It should be noted that in case of additive white Laplacian noise which models the impulse noise (Salt \& Pepper noise), we have the L1 norm data fidelity term~\cite{totalvariation}. Usually, L1 norm data term has better robustness against pixel outliers~\cite{BTV}. Without loss of generality, the data fidelity term can be formulated as 
\begin{linenomath}
\begin{equation}\label{eq:misr}
-log\left(\prod\limits_{i=1}^{k}P(\y_i|\x)\right) = \sum\limits_{i=1}^{k}||\A_i\x-\y_i||^p_p
\end{equation}
\end{linenomath}
with the $L_p$ norm $p\in\{1,2\}$.
%with the $L_p$ norm $1\leq p\leq 2$.

\begin{figure*}
	\centering
	 \includegraphics[scale=1.3]{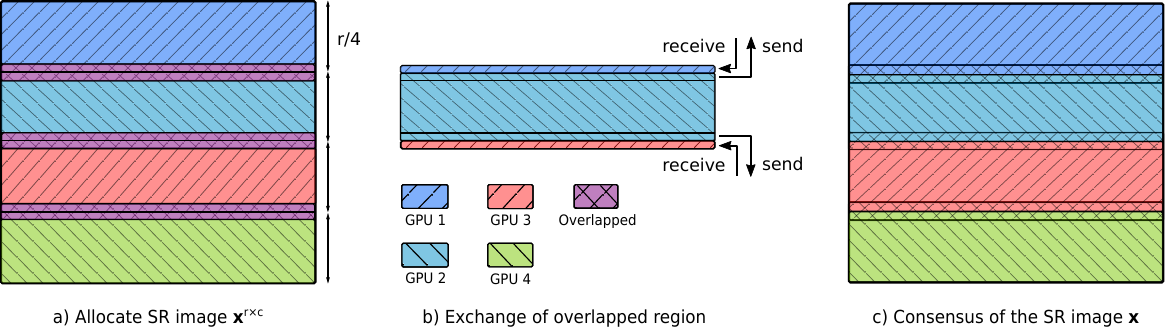}		
	\caption{Demonstration of the inner-outer border exchange scheme of the overlapped regions for 4 GPUs.}
	\label{fig:communication}
\end{figure*}

In the literature, there are several well-known handcrafted image priors $P(\x)$ including the total variation (TV)~\cite{TV}, Huber-Markov prior~\cite{Huber}, bilateral total variation (BTV)~\cite{BTV}, nonlocal total variation (NLTV)~\cite{NLTV}, and the more recent bilateral spectrum weighted total variation (BSWTV)~\cite{MPGBSWTV}. In this paper, aiming for reducing the computational complexity, we leverage the BTV as the image prior and the regularization term is therefore expressed as
\begin{linenomath}
\begin{equation}\label{eq:prior}
  -log(P(\x))= \sum\limits_{\mathbf{d}}
  \gamma(\mathbf{d})\parallel{\x-S_\mathbf{d}\x}\parallel_1,~ 
  \mathbf{d}=(d_x,d_y)
\end{equation}
\end{linenomath}
where $\mathbf{d}\in \mathbb{N}^2$ with $d_x, d_y\in[0,w-1]$ and $w$ denotes the window size accounting for the neighbors in the $x, y$ directions. $S_\mathbf{d}$ represents the shifting operator along $x$ and $y$ axis by $d_x$ and $d_y$ pixels. $\gamma(\mathbf{d}):=\alpha^{d_x+d_y}$ embodies the spatial decaying effect with constant $\alpha<1$. 

\begin{table}
\centering
\setlength{\tabcolsep}{5pt}
\captionsetup{justification=centering}
\caption{List of the consensus variables in SCG algorithm.}
\label{tab:parameters}
\begin{tabular}{V{4} c | l V{4}} 
 \hlineB{4}
Param. & Description \\ \hline
 $f_{c}, f_{c\_new}$  & consensus of the objective function \\  \hline
 %$f_{c\_new}$ & consensus of the updated objective function \\ \hline
 $\p_{c}$ & consensus of the conjugate weight vector  \\ \hline
 $\rr_{c}, \rr_{c\_new}$ & consensus of the steepest descent direction \\ \hline
 %$\rr_{c\_new}$ & consensus of the updated steepest descent direction \\ \hline
 $\sigma_{c}, \lambda_c$ & consensus of the scalars \\ \hline
 $\delta_{c}, \mu_{c}$ & consensus of the variables in step size \\ \hline
 %$\mu_{c}$ & consensus of the denominator of the step size \\ \hline
 $\alpha_{c}$ & consensus of the update step size\\
    \hlineB{3}     
	\end{tabular}
\end{table}

As the denominator $\prod\limits_{i=1}^{k}P(\y_i)$ of Eq.~\eqref{eq:bayes} is independent from the image $\x$, maximizing the posterior probability $P(\x|\y_1\dots\y_k)$ is equivalent to minimizing the negative logarithm of the numerator which is formulated respectively in Eqs~\eqref{eq:misr} and~\eqref{eq:prior}. Hence, we yield the overall objective function based on the MAP framework as following:
\begin{linenomath}
\begin{equation}\label{eq:objective}
  J(\x)= \sum\limits_{i=1}^{k}||\A_i\x-\y_i||^p_p + \lambda\sum\limits_{\mathbf{d}}
  \gamma(\mathbf{d})\parallel{\x-S_\mathbf{d}\x}\parallel_1
\end{equation}
\end{linenomath}
where the scaling factor of the fidelity term $\nicefrac{1}{2\sigma^2}$ in Eq.~\eqref{eq:log} is actually absorbed into the weighting parameter $\lambda$. In the experiments, we have used the L1 norm data term for a better robustness.

In order to accelerate the computation and alleviate the GPU memory load especially when coping with a sequence of large input images, we distribute the computational demand over multi-GPUs by data parallelism and follow a consensus-based convergence manner to guarantee a centralized solution. The latent SR image $\x$ is finally obtained by data fusion. %we decompose the objective function described in Eq.~\eqref{eq:objective} into subfunctions and each subfunction is solved by a single GPU.
In particular, Eq.~\eqref{eq:objective} can be rewritten as 
\begin{linenomath}
\begin{equation}\label{eq:reformulated}
  J(\x)= \sum\limits_{i=1}^{k} D_i(\x) + \lambda R(\x)
\end{equation}
\end{linenomath}
with $D_i$ representing the corresponding data term and $R$ being the regularization term. In this regard, the subfunction associated with the $h$th GPU is expressed as
\begin{linenomath}
\begin{equation}\label{eq:subfunction}
  J_h(\x_h)= \sum\limits_{i=1}^{k} D_i(\x_h) + \lambda R(\x_h),\quad s.t. ~\bigcup\limits_{h=1}^{g}\x_h=\x
\end{equation}
\end{linenomath}
where $\x_h$ is a fraction of the latent image $\x$ assigned to the $h$th GPU and $g$ denotes the number of employed GPUs. To enforce the distributed optimization towards a centralized solution, we allow communication between the local GPU node and the host CPU for a consensus update decision. Specially, we utilize the SCG algorithm~\cite{SCG} to iteratively solve the subproblem described in Eq.~\eqref{eq:subfunction} in each GPU. Instead of using the handcrafted step size or performing line search, SCG employs a step size scaling mechanism based on an adaptive scalar which achieves a faster and more robust convergence than the widely used approaches such as conjugate gradient with line search (CGL) and Broyden-Fletcher-Goldfarb-Shanno (BFGS). 

\begin{figure}[ht]
	\centering
	 \includegraphics[scale=0.8]{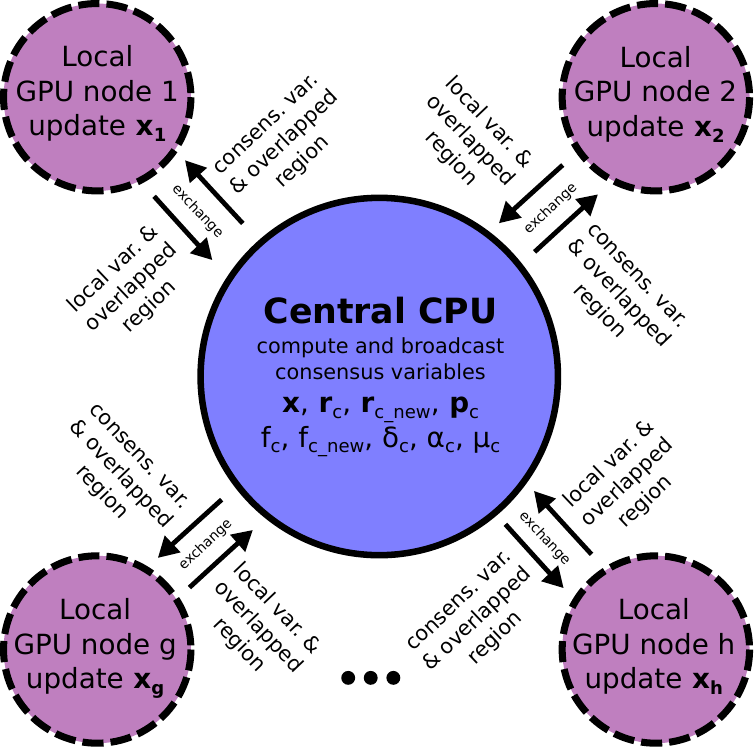}		
	\caption{Architecture of the proposed multi-GPU framework for MISR where $g$ GPU nodes are employed.}
	\label{fig:structure}
\end{figure} 

Aiming for synchronizing the update of the individual $\x_h$ towards a centralized solution, we unify the local SCG scalar variables $\sigma, \lambda, \delta, \mu, \alpha$ by data communication. As these variables are calculated based on the inner product of vectors, we can obtain the consensus variables by the aggregate of the broadcast local ones. By means of consensus variables, the subfunctions can converge synchronically and a homogeneous resolution among multi-GPUs is guaranteed. In Table~\ref{tab:parameters}, we list the unified scalar variables and vectors (in bold) of SCG. 

%Although a consensus convergence rate among different partitions is guaranteed by the exchange of the local SCG variables, since the distributed objective function in Eq.~\ref{eq:subfunction} contains the individual system matrix $\A$ and the TV regularization, both require the pixels in the overlapped region when computing the border pixels. 
In addition, to avoid border discontinuity of neighboring partitions, region overlapping between neighboring GPUs is required. Instead of the naive averaging of the overlapped regions which sacrifices the sharpness and visual quality, we perform an inner-outer border exchange in each SCG iteration as shown in Fig.~\ref{fig:communication}. A 4-GPU system is demonstrated and each GPU deals with the allocated image partition $\x_h$. The overlapped regions marked in violet are exchanged between neighboring GPUs. Particularly, since the inner borders can be correctly calculated only in case that the outer borders are consistent with the neighboring GPUs, the outer borders are replaced by the received ones and the inner borders are broadcast to the neighbors as exhibited in Fig.~\ref{fig:communication}b). Consequently, an agreement in the overlapped regions is achieved as shown in Fig.~\ref{fig:communication}c) without compromising the image sharpness. Without loss of generality, assuming $g$ GPU nodes are employed, the architecture of the proposed multi-GPU framework for SR is illustrated in Fig.~\ref{fig:structure}. The local variables and overlapped regions are interchanged in each SCG iteration over the host CPU and updated in a consensus scheme.

\begin{algorithm}
%\SetKwInOut{KwIn}{Local}
%\SetKwInOut{KwOut}{Central}
\caption{Distributed SR Reconstruction}\label{alg:pseudocode}
\begin{algorithmic}[1]
\State \text{Partition the LR images $\y_i, i\in[1\dots k]$ for each GPU} \newline
{node $h\in [1\dots g]$.}
\State \text{Initialize each GPU node with $\gamma(\mathbf{d}), \lambda, f_h, f_c, \p_h, \rr_h, \delta_c,$} \newline {$\mu_c, \alpha_c, \sigma, n_{iter}$.}
\State \text{Calculate matrices $\A_i, \A^T_i, i\in[1\dots k]$ in each GPU.}
\Procedure{Estimate latent image $\x$ according to Eqs.~\eqref{eq:objective} and~\eqref{eq:subfunction} using SCG~\cite{SCG}}{}
\While{$ i_{iter} < n_{iter}$}
\State \text{\textcolor{purple}{Local}\hspace{0.87em}: Calculate $||\p_h||^2_2, h\in [1\dots g]$.}
\State \text{\textcolor{blue}{Central}: Update $\sigma_c=\sigma/|\p_{c}|$, $||\p_{c}||^2_2 = \sum^g_h ||\p_h||^2_2$.}
\State \text{\textcolor{purple}{Local}\hspace{0.87em}: Calculate $\x_{h\_tmp}=\x_h+\sigma_c\p_h$.}
\State \text{\textcolor{blue}{Central}: Exchange overlapped regions of $\x_{h\_tmp}$}\par
\hspace{4.9em}{with neighboring GPUs.}
\State \text{\textcolor{purple}{Local}\hspace{0.87em}: Calculate $\delta_h$ according to SCG.}
\State \text{\textcolor{blue}{Central}: Update $\delta_c=\sum^g_h\delta_h$.}
\State \text{\textcolor{purple}{Local}\hspace{0.87em}: Calculate $\mu_h, \alpha_h$ according to SCG.}
\State \text{\textcolor{blue}{Central}: Update $\mu_c=\sum^g_h\mu_h, \alpha_c=\sum^g_h\alpha_h$.}
\State \text{\textcolor{purple}{Local}\hspace{0.87em}: Calculate $\x_{h\_new}=\x_h+\alpha_c\p_h$.}
\State \text{\textcolor{blue}{Central}: Exchange overlapped regions of $\x_{h\_new}$}\par\hspace{4.9em}{with neighboring GPUs.}
\State \text{\textcolor{purple}{Local}\hspace{0.87em}: Calculate $f_{h\_new}$ according to Eq.~\eqref{eq:subfunction}.}
\State \text{\textcolor{blue}{Central}: Update $f_{c\_new}=\sum^g_hf_{h\_new}$.}
\State \text{\textcolor{purple}{Local}\hspace{0.87em}: Calculate $||\rr_{h\_new}||^2_2$, inner product} \par \hspace{4.9em}{$\langle\rr_h,\rr_{h\_new}\rangle$.}
\State \text{\textcolor{blue}{Central}: Update $||\rr_{c\_new}||^2_2=\sum^g_h||\rr_{h\_new}||^2_2$,} \par \hspace{4.9em}{$\langle\rr_c,\rr_{c\_new}\rangle=\sum^g_h\langle\rr_h,\rr_{h\_new}\rangle$.} 
\State \text{\textcolor{purple}{Local}\hspace{0.87em}: Update $\p_h$.}%, boolen variable $isSuccess$.}
\State \text{\textcolor{blue}{Central}: $ i_{iter}=i_{iter}+1$.}
\EndWhile
\State \textbf{end while}
\State \text{\textcolor{blue}{Central}: Fuse $\x_h, h\in[1\dots g]$ to reconstruct $\x$.}
\State \textbf{return} \text{reconstructed image $\x$.}
\EndProcedure
\end{algorithmic}
\end{algorithm}

\begin{figure}
	\centering
	 \includegraphics[scale=0.95]{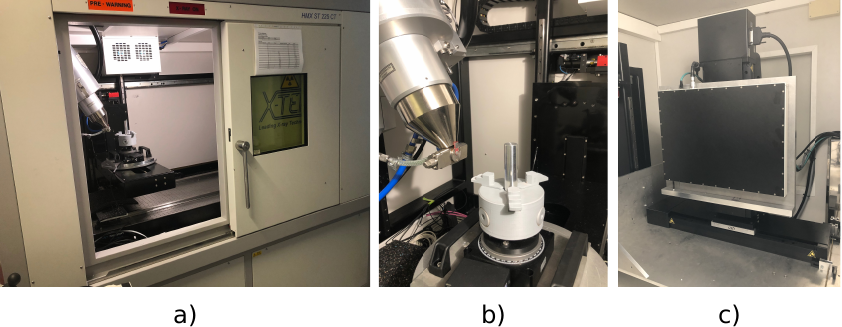}		
	\caption{CT scanner equipped with mounted linear stages. a) side view; b) 
X-ray tube and rotatable object (aluminium cylindrical phantom); (c) X-ray detector mounted on the controllable linear stages.}
	\label{fig:CT}
\end{figure}

In Algorithm~\ref{alg:pseudocode}, we present a detailed description of the proposed distributed optimization framework for MISR based on the SCG approach. The local GPU computation is marked by red and the centralized computation in the host CPU is denoted in blue. The local variables, overlapped regions, and consensus variables are respectively exchanged after the local and central update. The algorithm variables are initialized based on SCG~\cite{SCG} and the calculation of the system matrix $A_i$ is explained in Section~\ref{sec:Experiment}. The SR image $\x$ is fused when the SCG iterations are complete. 

\begin{figure*}
	\centering
	 \includegraphics[scale=0.95]{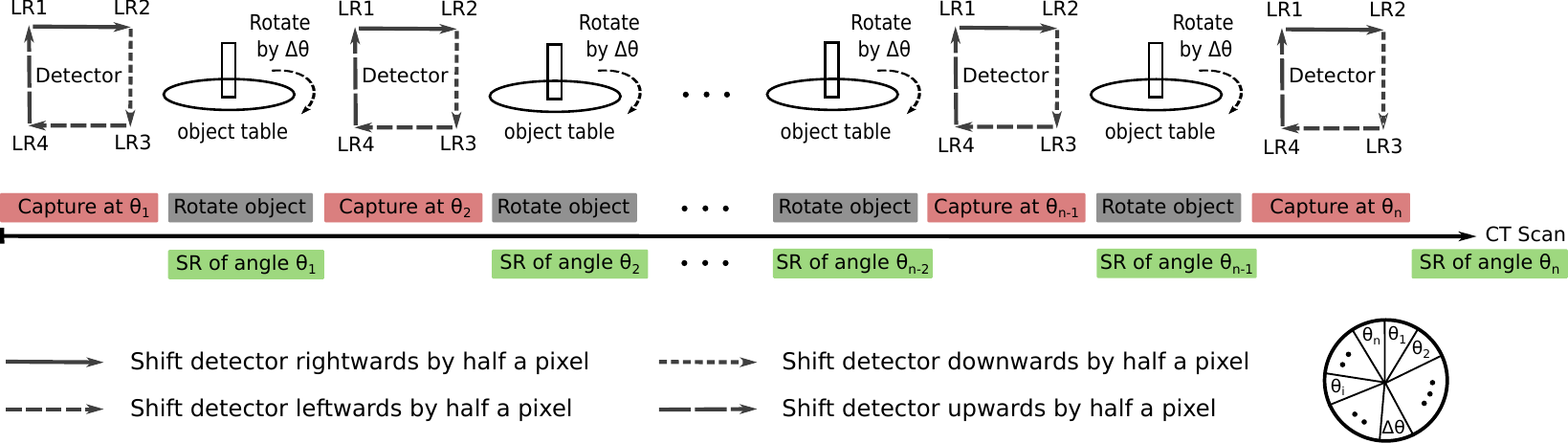}		
	\caption{Schematic illustration of the application of FL-MISR in CT imaging based on the controlled subpixel detector shift.}
	\label{fig:capture}
\end{figure*}

In the implementation, we have used the OpenCL framework. In order to optimize the data deployment on GPU memory, we exploited the local memory in the kernel functions to the most extent. Sparse matrix was employed to calculate the system matrix $\A_i=\D_i\B_i\M_i$ and the transpose $\A^T_i$ due to the sparseness of the downsampling, blurring, and motion matrices. Although memory transfer of local variables and overlapped regions between the GPU and host CPU is intended to hold the consensus convergence, transfer of large amounts of data is obviated during the SR reconstruction. %It is worthy noting that the overlapped regions are only required for the regularization term and the local update of $\x_i$ does not cover the outer border.  
It is worthy noting that the proposed distributed optimization framework is based on data parallelism and consensus SCG. It can be easily applied to other applications such as SISR and image denoising by replacing the objective function in Eq.~\ref{eq:objective}. %which is not confined to a specified objective function or a certain application. 

\subsection{Real-Time MISR for CT}

SR is always preferable in CT imaging where spatial resolution plays a determinant role in image quality assessment. We have applied the proposed FL-MISR on the industrial CT scanner as shown in Fig.~\ref{fig:CT}. %Particularly, the detector is mounted on the controllable linear stages for x- and y-positioning so that the detector can be shifted precisely to a predefined position. 
During the CT acquisition, the object is rotated by 360$^{\circ}$ and at each rotation angle, four LR projections (X-ray images) are captured via detector shift rightwards, downwards, leftwards, and upwards by half a pixel as illustrated in Fig.~\ref{fig:capture}.  %It should be noted that the presented approach is a generalized framework for MISR reconstruction based on multi-GPU system and is not confined to CT imaging. In order to enhance the spatial resolution of the reconstructed CT volume, we perform MISR on the captured projections of the same view . 
As long as all the four LR projections of the same view are collected, SR reconstruction is launched as denoted in green along the scan time axis. The capture-reconstruct fashion repeats until the whole CT acquisition is accomplished. Due to the fact that SR reconstruction usually takes less time than the accumulated time of projection acquisition (in red) and object rotation (in gray), SR can be performed in real-time during the CT scan without introducing extra runtime. The super-resolved projections are utilized for CT reconstruction and hence, an improved spatial resolution in CT is achieved by the increased detector sampling rate. We demonstrate the experimental results in Section~\ref{sec:Experiment}. It is necessary to note that since the same detector movement pattern is repeated for all the rotation angles during CT scan, the system matrices $A_i$ with $i\in[1,4]$ are calculated once at the beginning of the CT acquisition and shared by all the rotation angles. %for an upscaling of 3 or 4, 9 projections or 16 projections would be taken by shifting the detector by $\nicefrac{1}{3}$ or $\nicefrac{1}{4}$ pixel, respectively. %Comparing to the counterpart that performing SR on the reconstructed volumes at different detector positions, a major advantage of super-resolving X-ray images at each rotation angle is that the SR reconstruction can be performed in real-time along with the CT scan without introducing additional computation time. 

%\subsection{Multi-GPU Implementation}
%The proposed multi-GPU system was implemented using OpenCL. As the downsampling matrix $\D_i$, blurring matrix $\B_i$, and the motion matrix $\M_i$ are all sparse matrices, we used Eigen library~\cite{Eigen} to calculate the system matrix $\A_i=\D_i\B_i\M_i$ and the transpose $\A^T_i$. 

\section{Experiments and Results}
\label{sec:Experiment}
In this section, we conduct extensive experiments to evaluate the performance of the proposed FL-MISR from different aspects, mainly on resolution enhancement and computation acceleration. Specially, FL-MISR is evaluated for real-time CT imaging based on the synthetic and real-world CT measurements. Besides, the application of FL-MISR on natural images is evaluated using the public dataset DIV8K~\cite{DIV8K}. 

The CT measurements were carried out on the Nikon HMX ST 225 CT scanner as shown in Fig.~\ref{fig:CT} which is equipped with a flat panel Varian PaxScan 4030E detector of pixel size 127$\times$127 $\mu m$. The detector is mounted on the controllable linear stages for x- and y-positioning which supports detector displacement with a movement accuracy up to 1 $\mu m$. The focal spot size of the tungsten X-ray tube is power dependent and for the power under 7 $W$, which was utilized in our experiments, the effective focal spot size is about 6 $\mu m$ measured by the JIMA RT RC-04 micro chart. %As the spatial resolution in CT systems depends on the magnification of the measurement (the ratio between the source-detector distance and the source-object distance), we evaluated the effectiveness of FL-MISR at magnifications of 5, 10, and 25.

The calculation of the system matrix $\A_i$ is thoroughly described in our previous work~\cite{MPG}. For an upscaling of 2$\times$ with half pixel detector shift and a $3\times 3$ Gaussian blur for $B_i$, a 12-row block area in the HR grid is required as the overlapped region between neighboring GPUs. The weighting parameters $\lambda$ and $\alpha$ were respectively set as 0.05 and 0.4. The SCG iteration was limited to 20. In practice, larger $\lambda$ should be opted in case of strong noise and fewer SCG iterations should be used for fast CT acquisitions. %The SR reconstruction was performed on a cluster of Nvidia GeForce GTX 1080 GPUs with 11GB of RAM for each and the Intel Xeon Gold 6148 CPU equipped with 56 Cores and 755GB memory. 
To quantify the resolution enhancement by FL-MISR on CT systems, we adopted the modulation transfer function (MTF) which was measured according to the standard ASTM-E 1695.

\subsection{Evaluation of FL-MISR on Spatial Resolution Enhancement}
Before we evaluate FL-MISR on CT imaging, we briefly introduce the CT system and the assessment metric. CT scanner mainly consists of two components: the X-ray tube and the X-ray sensitive detector. The spatial resolution of the CT system is hence primarily limited by the focal spot size of the X-ray tube and the detector pixel size. Usually, spatial resolution of imaging systems is assessed by the MTF which is calculated as the normalized magnitude of the Fourier Transform of the point spread function (PSF). %describes the smallest visually distinguishable line pairs per $mm$.
The MTF of the CT system is formulated by $MTF_{sys}=MTF_{fs}\cdot MTF_{det}\cdot MTF_{others}$, where $MTF_{fs}$ and $MTF_{det}$ respectively denote the MTF of the X-ray focal spot and the detector. Other components such as the reconstruction algorithm, X-ray beam hardening, and display monitor are usually of less influence on the overall $MTF_{sys}$. %and are out of the scope of this paper. 
In this work, we perform subpixel detector shift to achieve a higher detector sampling rate which will lead to an effective improvement of $MTF_{sys}$ when $MTF_{det}$ dominates $MTF_{fs}$, which is usually the case in many CT applications. 

\begin{figure}
%\vspace{1em}
	\centering
	 \includegraphics[scale=0.97]{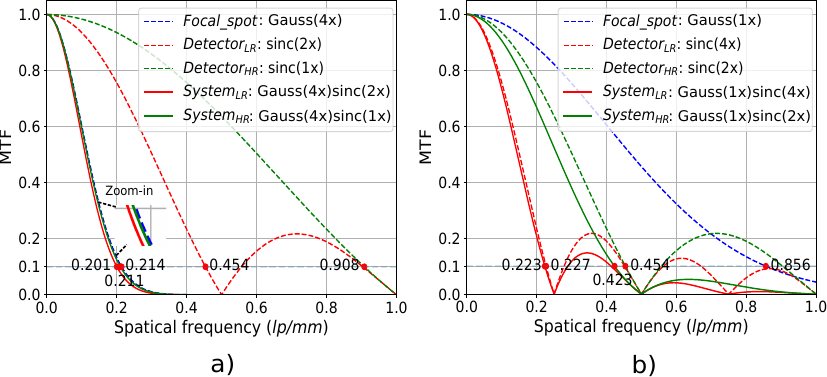}		
	\caption{Influence of improved detector MTF on the system MTF based on one-dimensional analysis. a) when $MTF_{fs}$ dominates, $MTF_{sys}$ rarely improves; b) in case of $MTF_{det}$ dominating, $MTF_{sys}$ improves significantly.}
	\label{fig:analysisMTF}
\end{figure}

\subsubsection{Evaluation on Synthetic CT Images}
In order to analyze the effectiveness of subpixel detector shift on the spatial resolution enhancement in CT, we firstly demonstrate the impact of $MTF_{det}$ on the $MTF_{sys}$. To simplify the system model, we consider only the primary components and therefore, we yield $MTF_{sys} := MTF_{fs}\cdot MTF_{det}$. The $MTF_{fs}$ is modeled by a Gaussian function and the $MTF_{det}$ is represented by a $sinc$ function due to the assumed rectangular shape of each pixel. As shown in Fig.~\ref{fig:analysisMTF}, the left plot indicates the case where $MTF_{fs}$ dominates $MTF_{det}$, for instance when the object is extremely close to the X-ray source and the right one depicts the situation where $MTF_{det}$ dominates. The MTF of the detector with full pixel size and with half pixel size is respectively denoted as $Detector_{LR}$ and $Detector_{HR}$. The MTF at $10\%$ is usually considered as the visible limit in practice and is marked by the gray dotted line. It is shown that halving the detector pixel size doubles the $MTF_{det}$ and improves the overall $MTF_{sys}$ effectively when $MTF_{det}$ dominates, while for the case $MTF_{fs}$ dominates, $MTF_{sys}$ has a negligible improvement. 

\begin{figure}
%\vspace{0.4em}
	\centering
	 \includegraphics[scale=0.8]{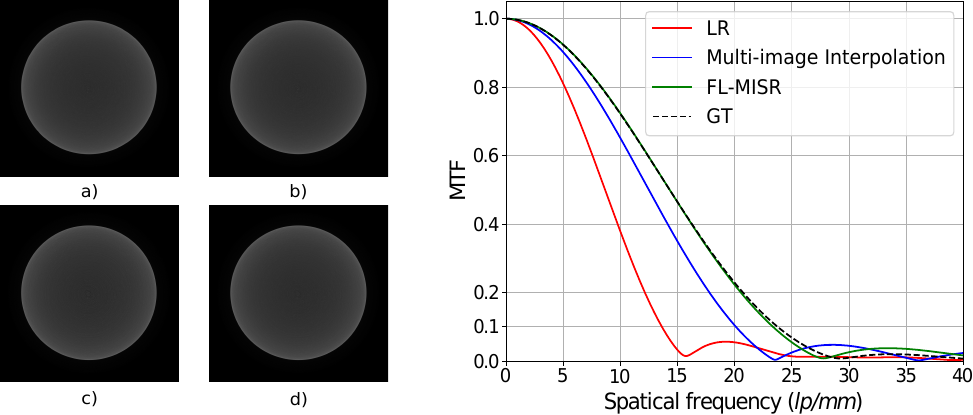}		
	\caption{Evaluation of MTF on the CT cross section of an aluminium cylindrical phantom. Left: a) LR, b) multi-image interpolation, c) FL-MISR, d) GT; Right: MTF.}
	\label{fig:SynMTFs}
\end{figure}

\begin{figure}
\vspace{0.1em}
	\centering
	 \includegraphics[scale=0.55]{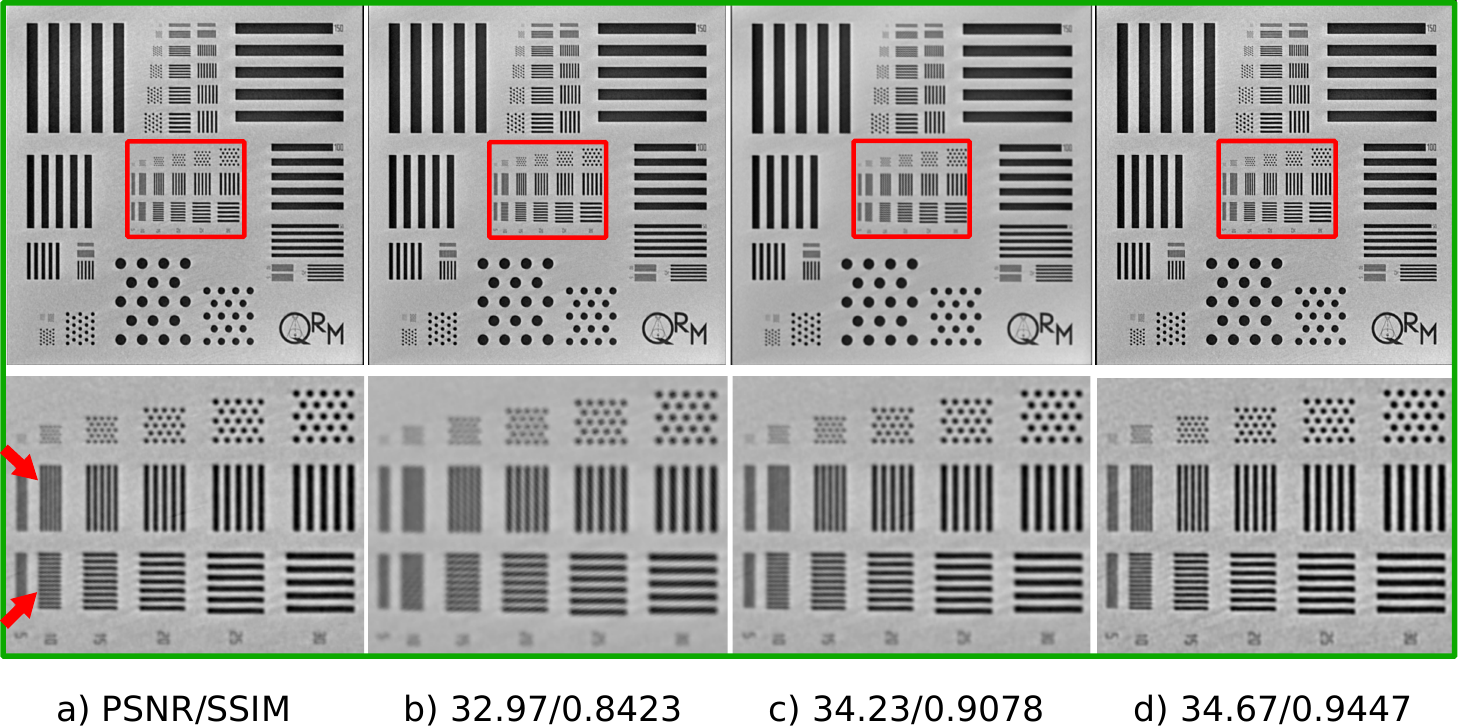}		
	\caption{CT images of the QRM bar pattern phantom. The ROIs are marked by red rectangle and zoomed in. a) GT; b) Bilinear interpolation; c) multi-image interpolation; d) FL-MISR.}
	\label{fig:analysisQRM}
\end{figure}

\begin{figure*}
	\centering
	 \includegraphics[scale=0.97]{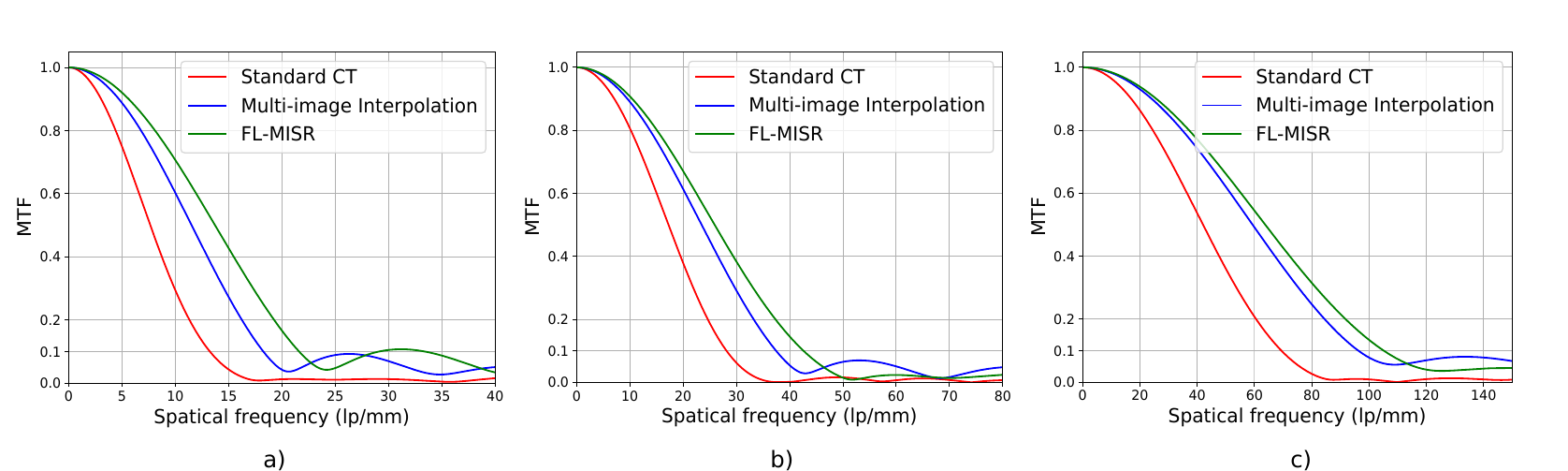}		
	\caption{Evaluation of MTF at different magnifications. a) magnification of 5; b) magnification of 10; c) magnification of 25.}
	\label{fig:MTFs}
\end{figure*}

\begin{figure*}
	\centering
	 \includegraphics[scale=0.82]{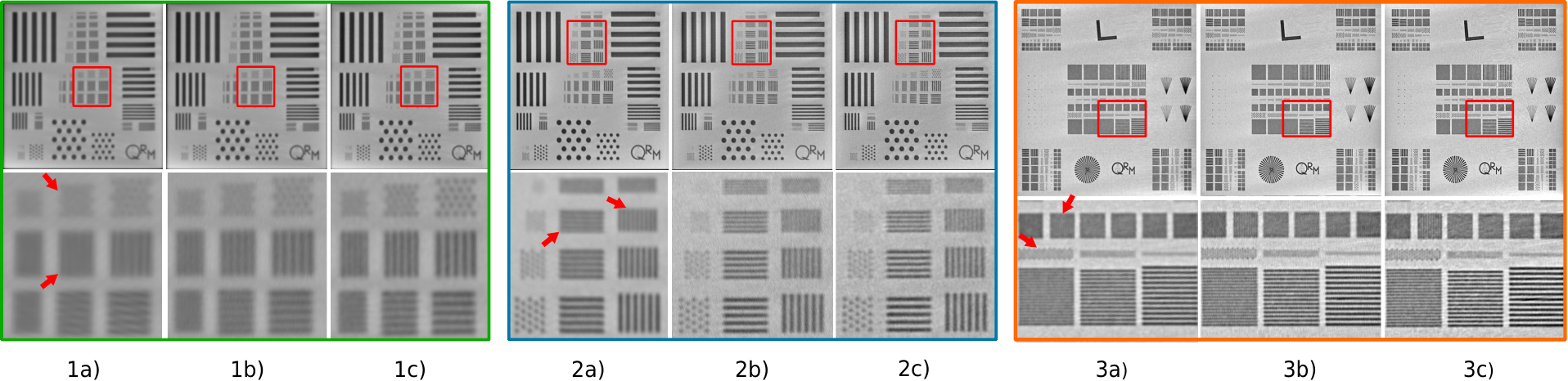}		
	\caption{CT images of QRM bar pattern phantom. Left (marked in green): magnification of 5; Middle (marked in blue): magnification of 10; Right (marked in orange): magnification of 25. a) standard CT without detector shift; b) multi-image interpolation; c) FL-MISR.}
	\label{fig:QRMMicro}
\end{figure*}

\begin{table*}[!b]
\centering
\setlength{\tabcolsep}{2pt}
\captionsetup{justification=centering}
\begin{threeparttable}
\caption{Parameter setup for CT measurements.}
\label{tab:setup}
\begin{tabular}{V{4} c | c | c | c | c | c | c | c V{4}} 
 \hlineB{4}
Test Phantoms & Voltage ($kV$) & Current ($\mu A$) & \# of Angles & Exposure ($s$) & Subpixel Shift & Magnification & Filter ($mm$)\\ \hline
Aluminium cylinder & 200 & 34 & 3600 & 3 & 0.5 & 5, 10, 25& Al 2.5\\  \hline
QRM bar pattern & 80 & 86& 3600 & 3 & 0.5  & 5, 10, 25& None\\  \hline
Dry concrete joint & 180 & 110 & 3600 & 3 & 0.5 & 3, 5 & Al 0.1\\
    \hlineB{3}     
	\end{tabular}
    \end{threeparttable}
\end{table*}

Based on the analysis above, we evaluate FL-MISR on the CT images quantitatively and qualitatively. Specially, we conducted CT scans of an aluminium cylindrical phantom with a diameter of 20 $mm$ as shown in Fig.~\ref{fig:CT}b) which was fixed perpendicular to the rotation table and a QRM bar pattern resolution phantom at the magnification of 20. Considering them as the ground truth (GT), we simulated four sets of 0.5$\times$ LR projections by shifting the GT projections rightwards, downwards, leftwards, and upwards by one pixel followed by a $2\times 2$ binning. The downscaled LR projections were fused by interpolation and by FL-MISR. As the inter-image offset is assumed to be one pixel and accurate, for interpolation-based fusion we inserted the pixel values of the LR images into the corresponding integer location in the HR grid. The super-resolved projections were then used for CT reconstruction by filter backprojection (FBP). The CT cross sections of the aluminium cylindrical phantom and the associated MTF are demonstrated in Fig.~\ref{fig:SynMTFs}. The LR CT was reconstructed by the reference (upper left) set of the downscaled projections. As we can clearly see that FL-MISR resembles the MTF of the GT extremely well and almost doubles the MTF of the LR image. %In order to analyse the robustness against noise, we contaminated the LR image by additive white Gaussian noise (AWGN) with $\sigma=500$. The associated MTF is depicted in Fig.~\ref{fig:SynMTFs}~b). We can observe that the MTF of multi-image interpolation degrades in the presence of noise, while FL-MISR is almost not influenced. 
To illustrate the performance of FL-MISR visually, we %generated the LR images of the QRM target following the same scenario as the aluminium cylindric phantom. 
present the CT images of the QRM bar pattern target in Fig.~\ref{fig:analysisQRM}. It is shown that FL-MISR provides a more pleasant result with sharper structures and better visual quality.

\iffalse
\begin{figure}
%\hspace{1em}
	\centering
	 \includegraphics[scale=0.66]{Fig10.pdf}		
	\caption{CT images of QRM bar pattern nano phantom at magnification of 25. a) standard CT without detector shift; b) multi-image interpolation; c) FL-MISR.}
	\label{fig:QRMNano}
\end{figure}
\fi
\begin{figure*}
%\hspace{-1em}
\centering
\includegraphics[scale=0.90]{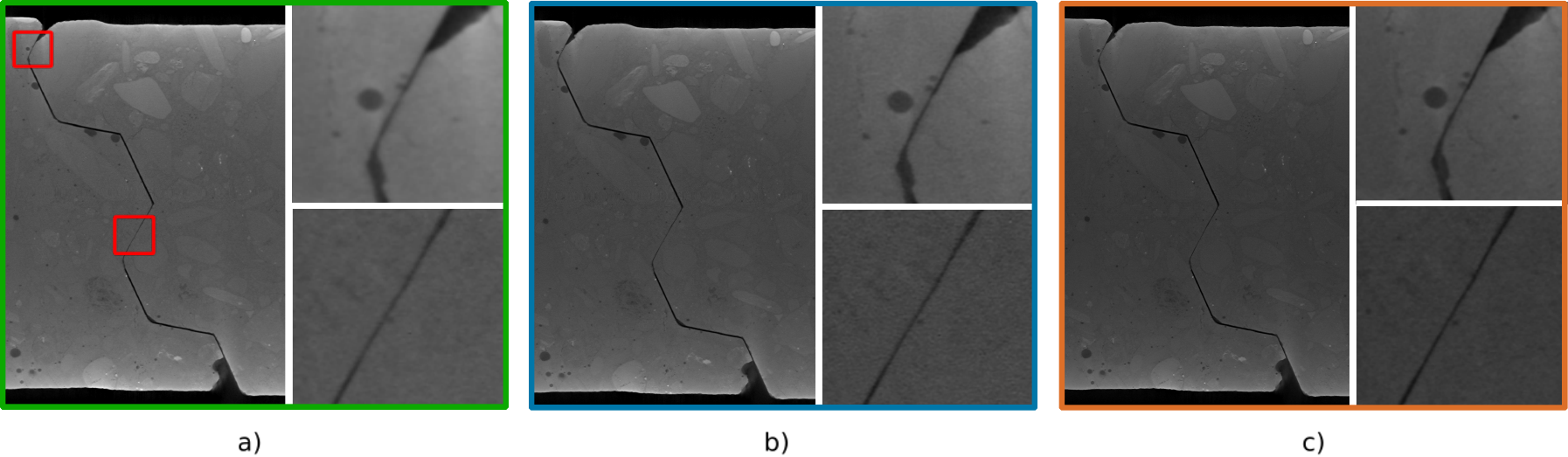}
\caption{CT images of a dry concrete joint with the ROI in the closeup views. a) standard CT without detector shift at the magnification of 3; b) FL-MISR with an upscaling of 2$\times$ at the magnification of 3; c) standard CT without detector shift at the magnification of 5.}
\label{fig:Concrete}
\end{figure*}

\subsubsection{Evaluation on Real-World CT Images}
As the spatial resolution of CT systems depends on the magnification, we evaluate FL-MISR on the real-world CT scans at different magnifications. Particularly, we conducted CT measurements of aluminium cylindrical phantoms with diameters of 10 $mm$ and 20 $mm$, QRM bar pattern phantom with spatial resolution ranging from 3.3 $lp/mm$ to 100 $lp/mm$, QRM bar pattern nano phantom which covers resolution from 50 $lp/mm$ to 500 $lp/mm$, and a cylindrical dry concrete joint with a diameter of 50 $mm$. The aluminium cylindrical phantoms and the QRM resolution targets were both scanned at magnifications of 5 (voxel size of 25.4 $\mu m$), 10 (voxel size of 12.7 $\mu m$), and 25 (voxel size of 5.08 $\mu m$) and the concrete joint was acquired at magnifications of 3 (voxel size of 42.3 $\mu m$) and 5. The detailed measurement setup is summarized in Table~\ref{tab:setup}. As illustrated in Fig.~\ref{fig:capture}, the X-ray detector was repeatedly displaced clockwise by half a pixel in a precisely controlled way. The projection at each detector position took 3 $s$, namely at each rotation angle 4$\times$3 $s$ was required for the acquisition. The object table rotated over 360$^\circ$ with 0.1 degree resolution following a stop-move manner and hence in total 4$\times$3600 projections were taken. Aluminium filters were utilized to absorb the soft X-ray beam and suppress the beam hardening artifact. We compare FL-MISR with multi-image interpolation and the standard CT without detector shift where the exposure time was set as 12 $s$, the same as FL-MISR. 

In Fig.~\ref{fig:MTFs}, we demonstrate the MTF measured by the aluminium cylindrical phantoms at different magnifications according to the standard ASTM-E 1695. It is shown that FL-MISR performs significantly better than the standard CT at all the investigated magnifications covering voxel size up to 5.08 $\mu m$. The multi-image interpolation behaves worse than FL-MISR as expected due to the naive manner of fusion. 

The CT images of the QRM bar pattern phantom and QRM bar pattern nano phantom are illustrated in Fig.~\ref{fig:QRMMicro} with the corresponding closeup views. Comparing to the standard CT images, we can observe that FL-MISR and multi-image interpolation both improve the spatial resolution by exploiting the additional information captured via subpixel detector shift. However, multi-image interpolation is less robust than the optimization-based FL-MISR. FL-MISR generates sharper edges  and provides more pleasant results in visual perception. In fact, the spatial resolution estimated by the visibility of the QRM bar patterns %according to the specifications of the QRM targets 
coincides with the MTF measured by the cylindrical phantoms.       

In Fig.~\ref{fig:Concrete}, we illustrate the CT images of a dry concrete joint with the zoomed-in region of interest (ROI). Fig~\ref{fig:Concrete}a and Fig~\ref{fig:Concrete}b represent respectively the results of the standard CT without detector shift and FL-MISR at the magnification of 3. Fig~\ref{fig:Concrete}c exhibits the results of standard CT at the magnification of 5 which is considered as the reference image. It is shown that comparing to the standard CT with a voxel size of 42.3 $\mu m$ at the magnification of 3, FL-MISR generates sharper contours with more detailed structures which resembles the CT measurement at the magnification of 5 better. % with a voxel size of 25.4 $\mu m$ better. %Some texts here xxx xxx xxx for the improvement of surface determination/segmentation.

\begin{figure}
	\centering
	 \includegraphics[scale=0.62]{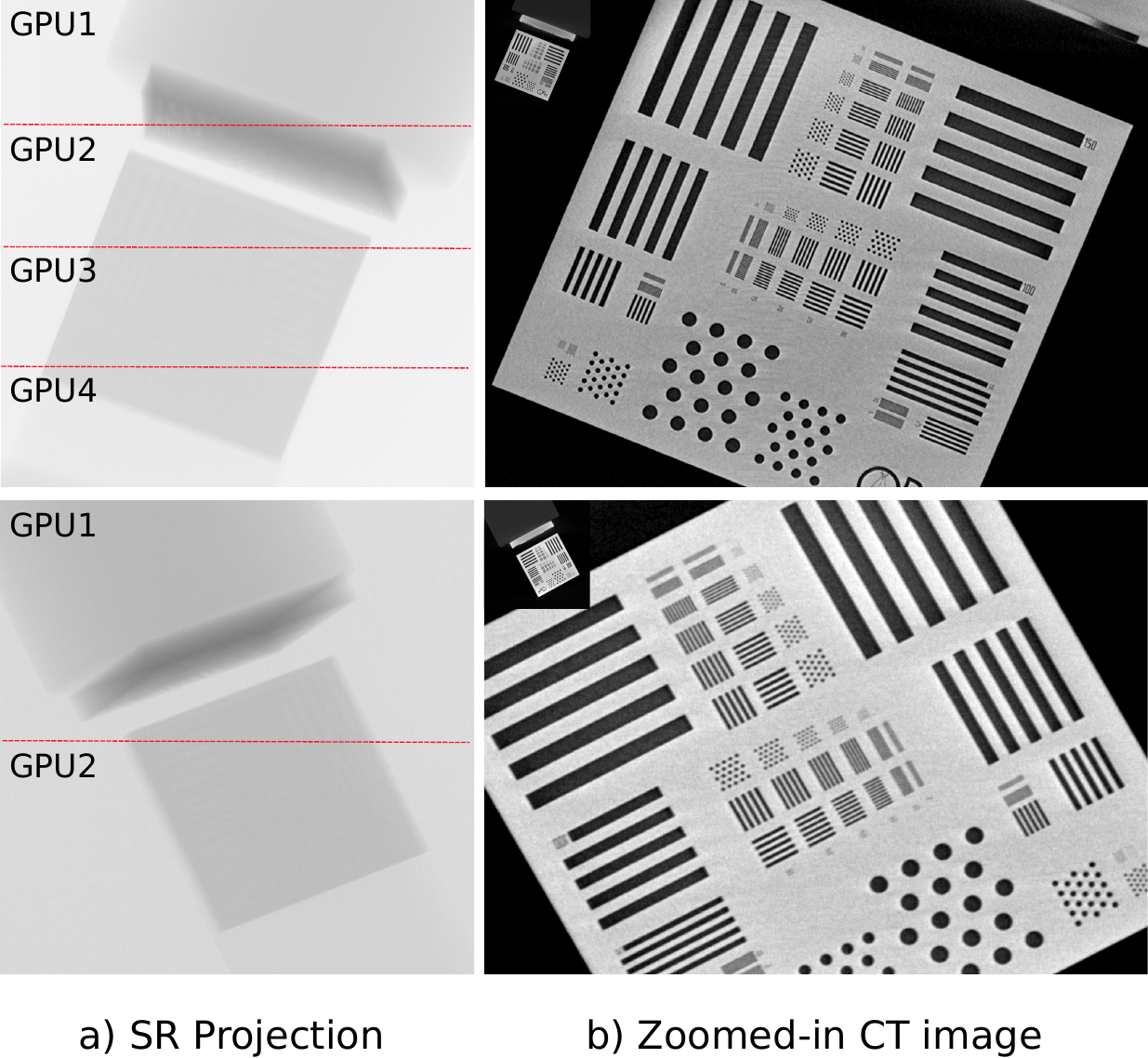}		
	\caption{Evaluation on the border effect. First row: on the synthetic volume as utilized in Fig.~\ref{fig:analysisQRM}; Second row: on the real-world volume as used in the middle graph of Fig.~\ref{fig:QRMMicro}. Red dotted line marks out the border of the partitions allocated to the GPUs.}
	\label{fig:Bordereffect}
\end{figure}

\begin{table*}[!b]
\centering
\setlength{\tabcolsep}{2pt}
\captionsetup{justification=centering}
\begin{threeparttable}
\caption{Evaluation of FL-MISR on 8-bit natural images in DIV8K dataset. MI Interp.: Multi-image interpolation.}
\label{tab:natural}
\begin{tabular}{V{4} c | c | c | c | c | c | c | c | c V{4}} 
 \hlineB{4}
 \multicolumn{2}{V{4}c|}{Image Index} & \#0001 &\#0002 & \#0007 & \#0027 & \#0055 & \#0066 & \#0084 \\ \hline
\multicolumn{2}{V{4}c|}{Resolution of GT} &  5376$\times$5760 & 5568$\times$5760 & 1920$\times$2880 & 2112$\times$2880 & 5760$\times$5760 & 1920$\times$2880 & 5760$\times$3840\\  \hline
\multicolumn{9}{V{4}cV{4}}{Upscaling 2$\times$} \\  \hline
\multirow{2}{*}{MI Interp.} &PSNR/SSIM &30.49/0.9215 &28.44/0.8677 &33.68/0.8810 &28.37/0.8988 &33.80/0.9018 & 35.21/0.9296 &29.11/0.8277\\  
& Runtime ($s$) & 0.51&0.52 &0.11 &0.20 &0.53 &0.11 &0.36\\ \hline
\multirow{2}{*}{FL-MISR} &PSNR/SSIM &37.11/0.9620 &32.99/0.9360 &35.09/0.9111 &33.21/0.9417 &38.03/0.9564 & 37.12/0.9452 &34.13/0.9410\\  
& Runtime ($s$) &1.50 &1.29 &0.69 &0.71 &1.3 &0.66&1.21\\ \hline
\multicolumn{9}{V{4}cV{4}}{Upscaling 3$\times$} \\  \hline
\multirow{2}{*}{MI Interp.} &PSNR/SSIM &26.74/0.8460 &25.65/0.7749 &32.03/0.8395 &25.15/0.8212 &30.79/0.8153 &32.65/0.8968 &26.19/0.6883 \\  
& Runtime ($s$)&1.00 &0.99 &0.11 &0.13 &0.55 & 0.11&0.38\\ \hline
\multirow{2}{*}{FL-MISR} &PSNR/SSIM &33.24/0.9446 &29.43/0.8941 &33.99/0.8941 &30.17/0.9139 &35.90/0.9379 &36.06/0.9398 &30.54/0.8764 \\  
& Runtime ($s$)&1.78 &1.73 &0.32 &0.38 &1.93 &0.35&1.65\\ \hline
    \hlineB{3}     
	\end{tabular}
    \end{threeparttable}
\end{table*}

\subsubsection{Evaluation on Border Effect and Consensus Convergence}
\begin{figure}
	\centering
	 \includegraphics[scale=0.9]{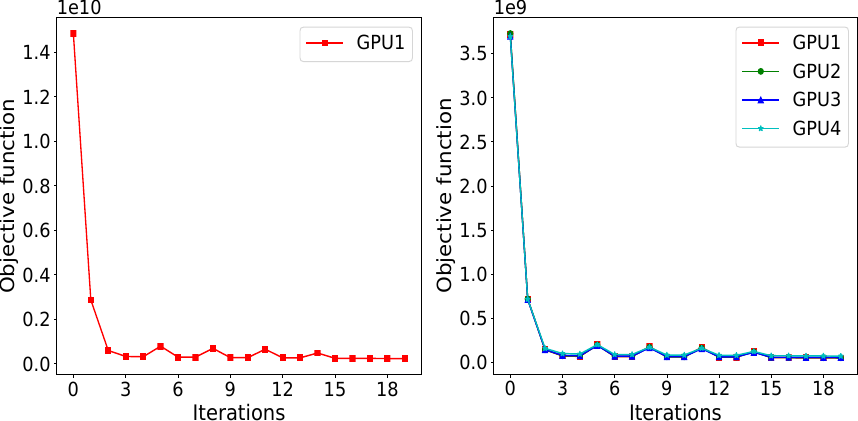}		
	\caption{Evaluation on consensus convergence based on the objective function. Left: convergence curve obtained using single GPU; Right: convergence curves obtained using 4 GPUs.}
	\label{fig:Consensus}
\end{figure}

As explained in Fig.~\ref{fig:communication}, we exchange the overlapped regions between neighboring GPUs to avoid border discontinuity. In Fig~\ref{fig:Bordereffect}, we demonstrate the super-resolved projections and the associated CT images of the synthetic (top row) and the real-world measurements (bottom row). For the synthetic image, we employed four GPUs and for the real-world one, two GPUs were in use. The individual $\x_h$ of each GPU is partitioned by the red dotted line. As we can observe that the overlapped regions, a 12-row block surrounding the borders (the red dotted lines), are of inherent sharpness without intensity discontinuity and the border effect is fundamentally obviated. Besides, in order to avoid inhomogeneous resolution in different partitions, we synchronize the update of the partitioned $\x_h$ among all the GPUs by exchanging the local variables of SCG. In Fig.~\ref{fig:Consensus}, we illustrate the convergence curve of the centralized objective of Eq.~\ref{eq:objective} running on a single GPU and the distributed objective of Eq.~\ref{eq:subfunction} running on 4 GPUs. The consensus convergence is reflected in two aspects. First, the 4 GPUs have exactly the same convergence trend, where they are almost overlaid, due to the share of the SCG variables. Second, the distributed objective follows the same convergence trend as the centralized one and moreover, the sum of the 4 distributed objectives equals the centralized one by resorting to the scheme we adopt for the calculation of the consensus variables of SCG as described in Section~\ref{subsec:distributed optimization}. In addition, we can observe that the objective function is almost converged after 5 SCG iterations.    

\begin{figure}
	\centering
	 \includegraphics[scale=0.95]{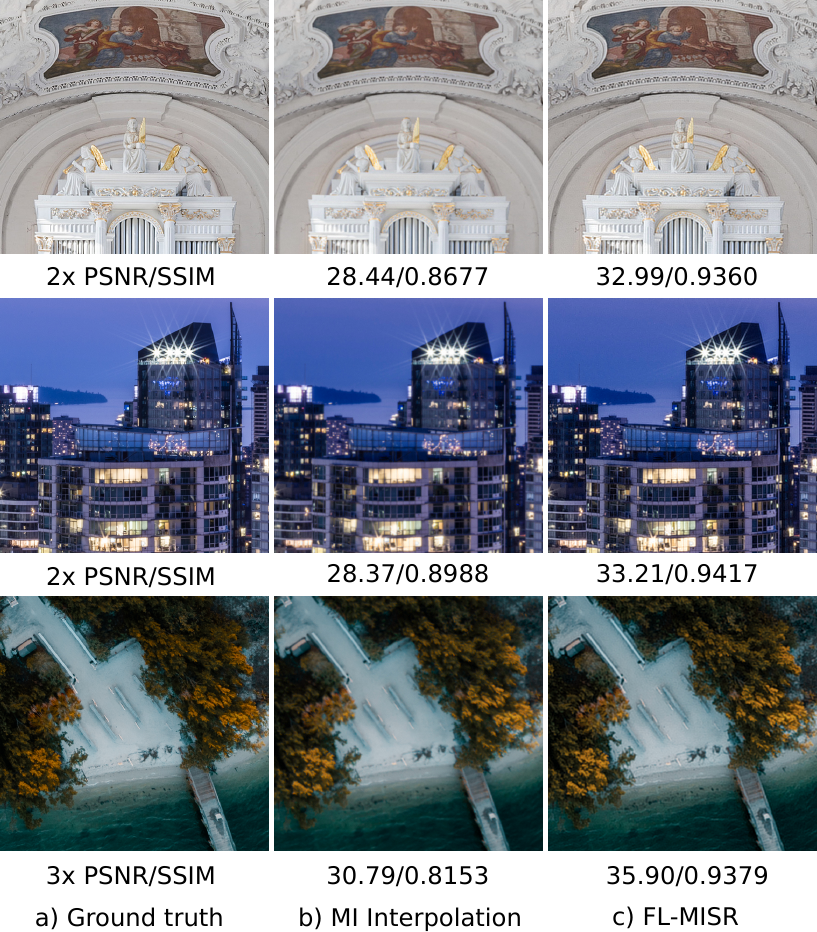}		
	\caption{Evaluation on natural image dataset DIV8K. a) GT; b) Multi-image interpolation; c) FL-MISR. The first two rows are for the upscaling of 2$\times$ and the bottom row is for the upscaling of 3$\times$ (Better viewed in color).}
	\label{fig:natural}
\end{figure}

\begin{table*}
\centering
\setlength{\tabcolsep}{3pt}
\captionsetup{justification=centering}
\begin{threeparttable}
\caption{Evaluation of computation time in terms of input image size, number of SCG iterations, and CPU/GPU platforms for the upscaling of 2$\times$ where four 16-bit input images were utilized. (N/A indicates not applicable.)}
\label{tab:runtime}
\begin{tabular}{V{4} c | c c c | c c c | c c c | c c c | c c c V{4}} 
 \hlineB{4}
Input image size & \multicolumn{3}{c|}{512$\times$512} & \multicolumn{3}{c|}{1024$\times$1024} &\multicolumn{3}{c|}{2048$\times$2048} & \multicolumn{3}{c|}{2300$\times$3200} & \multicolumn{3}{c V{4}}{4096$\times$4096} \\ \hline
\multicolumn{16}{V{4}cV{4}}{Non-iterative method} \\ \hline
Multi-image interp. & \multicolumn{3}{c|}{0.03} & \multicolumn{3}{c|}{0.07} &\multicolumn{3}{c|}{0.26} & \multicolumn{3}{c|}{0.45} & \multicolumn{3}{c V{4}}{2.07} \\ \hline
\multicolumn{16}{V{4}cV{4}}{Proposed iterative FL-MISR} \\ \hline
SCG iterations & 5 & 10 & 20 &  5 & 10 & 20 &5  & 10 & 20 &5  & 10 & 20 &5  & 10 & 20 \\  \hline
CPU$^*$ ($s$) & 1.06 & 2.24& 4.48&  4.08& 8.64 & 17.37 &16.21& 34.89 &69.02 &23.86& 50.68 &113.96 &49.67& 105.71 &250.82 \\  \hline
1 GPU ($s$) & 0.08 & 0.13& 0.25&  0.22& 0.42 & 0.78 &0.70& 1.30 &2.43 &N/A& N/A &N/A &N/A& N/A &N/A \\  \hline
%2 GPU (sec) & 0.11 & 0.21& 0.39&  0.21& 0.37 & 0.67 &0.56& 0.90 &1.58 &0.89& 1.47 &2.64 &--& -- &-- \\  \hline
4 GPU ($s$) & 0.07 & 0.12& 0.22&  0.25& 0.44 & 0.79 &0.52& 0.76 &1.32 &0.79& 1.20 &2.33 &2.38& 3.02 &4.33\\ 
    \hlineB{3}     
	\end{tabular}
	\begin{tablenotes}
   	\footnotesize
   	\textsuperscript{*}CPU experiments were conducted on the Intel Xeon Gold 5120 CPU equipped with 56 cores.
    \end{tablenotes}
    \end{threeparttable}
\end{table*}

\subsubsection{Evaluation on Natural Images}
Since the distributed optimization of FL-MISR is based on data parallelism, FL-MISR is not limited to a certain application. We evaluate the proposed FL-MISR on natural images using the public dataset DIV8K~\cite{DIV8K}. Particularly, we randomly selected 7 natural images with the vertical or horizontal resolution ranging from 1920 to 5760 pixels as the GT. For each GT image, 4 and 9 LR images were respectively generated for upscaling factors of 2$\times$ and 3$\times$ according to Eq.~\ref{eq:sisr} with $\varepsilon\sim N(0,1)$ and translational movement of $\nicefrac{1}{2}$ and $\nicefrac{1}{3}$ pixel. We performed SR reconstruction only for the luminance channel on 4 GPUs and set the SCG iterations as 10. The SR performance is assessed by PSNR, SSIM, and runtime. Quantitative evaluation is summarized in Table~\ref{tab:natural}. As we can see, the proposed FL-MISR outperforms the multi-image interpolation by a large margin in PSNR and SSIM. Although the iterative FL-MISR requires 2$\sim$5$\times$ runtime as the naive interpolation one, it supports an SR output of $5760\times5760$ resolution within 1.3$s$ for the upscaling of 2$\times$ and 1.93$s$ for the upscaling of 3$\times$. It is interesting to find that the runtime of SCG depends not only on the image size but also on the count of iterations with successful reduction in the objective function as expressed in~\cite{SCG}. In Fig.~\ref{fig:natural}, we illustrate the closeup views of images $\textit{0002}, \textit{0027}$, and $\textit{0055}$ of DIV8K. The top two rows demonstrate the results for an upscaling of 2$\times$ and the bottom row is for the upscaling of 3$\times$. We can observe that FL-MISR provides pleasant results with significantly better visual quality than the multi-image interpolation. 

\begin{figure}
	\centering
	 \includegraphics[scale=1]{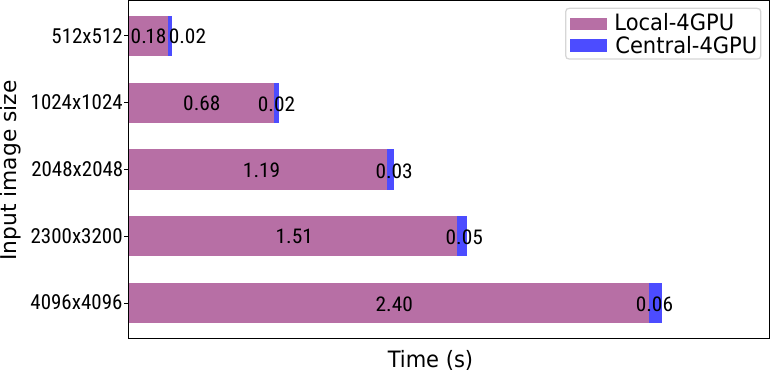}		
	\caption{Runtime distribution for the local and centralized computation for super-resolving images of different sizes by an upscaling of 2$\times$ under 20 SCG iterations on 4 GPUs.}
	\label{fig:GPUProfile}
\end{figure}

\subsection{Evaluation of FL-MISR on Acceleration}
In order to demonstrate the performance of FL-MISR in acceleration, we conducted SR reconstruction of different sized inputs ranging from 512$\times$512 to 4096$\times$4096 for an upscaling factor of 2$\times$ on a multi-core CPU, single GPU, and multi-GPU systems. In particular, the CPU experiments were performed on the Intel Xeon Gold 5120 CPU with 755GB memory which contains two nodes and each is equipped with 28 cores. The GPU experiments were carried out on the Nvidia GeForce GTX 1080 GPUs with 11GB memory. Since FL-MISR is based on the iterative SCG algorithm, we evaluated the runtime also with regard to the number of SCG iterations. Besides, we also demonstrate the runtime of the multi-image interpolation as the baseline. The performance of different configurations was calculated based on an average of 100 runs and is summarized in Table~\ref{tab:runtime} where N/A denotes not applicable due to the large GPU memory footprint. As we can see, comparing to the 56-core CPU variant, the single GPU implementation accelerates the computation by more than 25$\times$ for LR images of size 2048$\times$2048 and the multi-GPU implementation which uses 4 GPUs achieves a speedup up to 50$\times$. For large-scale images of size 2300$\times$3200 and 4096$\times$4096, FL-MISR running on 4 GPUs obtains a more than 55$\times$ speedup than the CPU implementation, while single GPU can not fulfill the memory requirement. For small sized inputs like 512$\times$512 and 1024$\times$1024, single GPU implementation has similar performance as multi-GPU and achieves a 20$\times$ speedup comparing to the multi-core CPU. Although the iterative FL-MISR requires more runtime than the naive interpolation one, FL-MISR has much better SR performance and the runtime difference becomes less as the image dimension increases. 

In addition, we analyzed the runtime distribution for the local and central computation on a 4-GPU system where the data communication time is aggregated into the central computation. We exhibit the average runtime distribution over 100 runs for input images of different sizes in Fig.~\ref{fig:GPUProfile}. It is shown that the consumed time for consensus computing is almost negligible comparing to the local computation, while it is fundamentally necessary to avoid border effects between neighboring GPUs and guarantee a consensus convergence over multi-GPU systems.

\section{Conclusion}
In this paper, we propose a multi-GPU accelerated large-scale multi-image super-resolution (MISR) framework based on data parallelism. Specially, each GPU node accounts for a designated region of the latent high-resolution (HR) image by applying an adapted scaled conjugate gradient (SCG) algorithm to the distributed subproblem. The local variables of the SCG algorithm are broadcast and aggregated in each iteration to synchronize the convergence rate over multi-GPUs towards a centralized optimum and consistent resolution. Furthermore, an inner-outer border exchange mechanism is performed in the overlapped regions of neighboring GPUs to avoid border effect without compromising the sharpness.  %The local variables of the SCG algorithm and the overlapped regions between neighboring GPUs are broadcast and shared to ensure a consensus convergence over multi-GPUs and avoid border effects. 

The proposed FL-MISR is seamlessly integrated into the computed tomography (CT) systems by super-resolving projections of the same view captured via subpixel detector shift. The SR reconstruction is performed on the fly during the CT acquisition such that no additional computation time is induced. Extensive experiments were conducted based on simulated data and real-world CT measurements of cylindrical phantoms, QRM bar pattern resolution targets, and cylindrical dry concrete joints to quantitatively and qualitatively evaluate the proposed FL-MISR. Experimental results demonstrate that the spatial resolution of CT systems is significantly improved in modulation transfer function (MTF) and visual perception by the application of FL-MISR. % by leveraging the resolution-enhanced projections. 
Moreover, comparing to a multi-core CPU implementation, the multi-GPU accelerated FL-MISR achieves a more than 50$\times$ speedup on a 4-GPU system and it is shown that the exchange of local SCG variables and overlapped regions between GPUs has limited impact on the overall runtime. Last but not least, evaluation on public dataset DIV8K shows that FL-MISR is not confined to CT imaging but also provides very promising results for natural images.

\begin{acknowledgements}
This work was supported by the German Research Foundation (DFG, Germany) under the DFG-project SI 587/18-1 in the priority program SPP 2187.
\end{acknowledgements}

\section*{Conflict of interest}
The authors declare that they have no conflict of interest.

\bibliographystyle{unsrt}
\bibliography{bibliography}

\iffalse

\fi
\end{document}